\documentstyle[12pt]{article}

\begin{document}

\renewcommand{\thesection}{\arabic{section}.} 
\renewcommand{\theequation}{\thesection \arabic{equation}}
\newcommand{\scs}{\setcounter{equation}{0} \setcounter{section}}
\def\req#1{(\ref{#1})}
\newcommand{\be}{\begin{equation}} \newcommand{\ee}{\end{equation}} 
\newcommand{\ba}{\begin{eqnarray}} \newcommand{\ea}{\end{eqnarray}} 
\newcommand{\la}{\label} \newcommand{\nb}{\normalsize\bf} 
\newcommand{\lb}{\large\bf} \newcommand{\vol}{\hbox{Vol}}
\newcommand{\bb} {\bibitem} \newcommand{\np} {{\it Nucl. Phys. }} 
\newcommand{\pl} {{\it Phys. Lett. }} 
\newcommand{\pr} {{\it Phys. Rev. }} \newcommand{\mpl} {{\it Mod. Phys. Lett. }}
\newcommand{\sg}{{\sqrt g}} \newcommand{\sqhat}{{\sqrt{\hat g}}}
\newcommand{\sqphi}{{\sqrt{\hat g}} e^\phi} 
\newcommand{\sqalpha}{{\sqrt{\hat g}}e^{\alpha\phi}}
\newcommand{\tp}{\cos px\ e^{(p-{\sqrt2})\phi}} \newcommand{\stwo}{{\sqrt2}}
\newcommand{\tr}{\hbox{tr}}

\begin{titlepage}
\renewcommand{\thefootnote}{\fnsymbol{footnote}}

\hfill CERN--TH/99--406

\hfill hep-th/9912156

\vspace{.4truein}
\begin{center}
 {\LARGE $AdS_5$ and the $4d$ Cosmological Constant}
 \end{center}
\vspace{.7truein}

 \begin{center}

 Christof Schmidhuber\footnote{schmidhu@mail.cern.ch}

 \vskip5mm

 {\it CERN, Theory Division, 1211 Gen\`eve 23}

 \end{center}

\vspace{1truein}
\begin{abstract}

\noindent
The hypothesis is discussed that our universe is really 5--dimensional
with a nonzero cosmological constant that produces a large negative
curvature. In this scenario, the observable 
flat 4--dimensional world is identified with the holographic projection
of the 5--dimensional world onto its own boundary.

\vfill
\noindent
CERN--TH/99--406
 
\noindent
December 99

\end{abstract}
 \renewcommand{\thefootnote}{\arabic{footnote}}
 \setcounter{footnote}{0}
\end{titlepage}

 \noindent
 {\small\it ``Very often a source of strong poetry and strong science
 is a good metaphor. My favorite one is Plato's cave:
 the parable of men sitting in a dark cave, watching the moving
 shadows on its wall. They think that the shadows are ``real''
 and not just projections of the outside world. It seems to me that
 the latest stages of the ongoing struggle to understand
 interactions of elementary particles create a picture stunningly
 close to this parable.''}

{\hfill \small\it A. M. Polyakov in \cite{pol}}

\vskip10mm
\noindent
\subsubsection*{1. Motivation}\scs{1}

\vskip2mm
\noindent
It has recently been
 conjectured that various conformally invariant 4--dimensional
 gauge theories including
${\cal N}=4$ supersymmetric $SU(N)$ 
Yang Mills theory with `t Hooft coupling
$\lambda\ =\ g_{YM}^2N$ 
have a strong coupling description in terms of type IIB superstring theory
compactified on a 5-dimensional Einstein manifold $E^5$
\cite{mal, kle, kgp, wit1, klewit, gub, oz}. 
The dilaton is constant, $\Phi=\Phi_0$, and the string coupling constant
is related to the gauge coupling constant by
$\kappa=e^{\Phi_0} = g_{YM}^2\ .$
The 
Einstein manifold has constant curvature radius
$$L\ \sim\ \lambda^{{1\over4}}$$
and $N$ units of electric and magnetic Ramond-Ramond flux flow through it.
The geometry of the remaining 5-dimensional spacetime is that of
anti-de Sitter space $AdS_5$ with the same curvature radius $L$:
$$ds^2=d\phi^2+e^{-{2\over L}\phi}dx_{||}^2\ .$$
$x_{||}$ parametrizes the 4-dimensional
space in which the gauge theory lives, and $\phi$
can be regarded
as ``renormalization group time'' in the sense that
$4d$ scale transformations 
$$x_{||}\rightarrow x_{||}e^\tau$$
can be absorbed by the shift
$$\phi\rightarrow\phi+L\tau\ .$$
The AdS boundary at $\phi=-\infty$ represents the UV limit of the gauge theory
\cite{suss}.
$\phi$ is a new  physical  coordinate, e.g. in the sense that
various objects in the gauge theory,
such as instantons or ``Pfaffian particles'' in the case of 
gauge group $SO(N)$ are localized both in the $x_{||}$ and the $\phi$
directions \cite{wit.bary, gre}.  

String loop corrections are proportional to ${1\over N^2}$,
while $\alpha'$-corrections are proportional to
${1\over\sqrt\lambda}$.
In particular, the strong-coupling limit $\lambda\rightarrow\infty$
can, for large $N$, be investigated in the classical supergravity approximation.
It is hoped that non-supersymmetric Yang-Mills theory and perhaps QCD
 have a similar description in terms of a 
5-dimensional string theory on some different background \cite{wit2,gro.oog}.

One immediate question might be: why don't we see the fifth dimension?
If it is true that nonabelian gauge theory at large scales
(large enough for the coupling to be strong) is best described by a dual
string theory in a 5-dimensional curved background, 
why then does the real world -- which contains $SU(3)$ gauge theory --
look 4-dimensional and flat, rather than 5-dimensional
and curved? In the case of ${\cal N}=4$ SYM, the answer is:
 even though $AdS_5$ is strongly curved,
it has a flat 4--dimensional Minkowskian boundary, and at scales
much larger than the curvature radius $L$ the
5--dimensional world essentially reduces to its own projection onto this 
boundary.\footnote{This is similar to Kaluza-Klein reduction
on a circle of radius $L$, where at scales much larger than $L$
the Kaluza-Klein modes drop out and an effective $4d$ theory is seen \cite{rs}.
In our case the circle is replaced by the noncompact coordinate $\phi$,
so things are more subtle, but the conclusion comes out essentially the same.}

Given this answer, it is very tempting to
use it for                 
a seemingly unrelated question that can already be asked purely in the context
of Einstein gravity and a priori has nothing to do  with 
the AdS/CFT correspondence:
why does our 4--dimensional world look flat and large, given that there should naturally
be a huge cosmological constant which
should strongly curve space-time, with a tiny curvature radius?

That is, it seems tempting to
assume that our world is really 5-dimensional rather than 4-dimensional.
Let us further assume that the $5d$
 cosmological
constant $\Lambda_5$ 
is huge as it should be. But let us assume it comes out such that the
sign of the $5d$ curvature is
negative. So the $5d$ space becomes anti-de Sitter, with curvature radius
$$L^2\ \propto\ {1\over\kappa_5^2\Lambda_5}\ ,$$
and so it acquires a boundary ($\kappa_5^2$ is the $5d$ Newton constant). 
Although the curvature radius of $AdS_5$ is small,
its $4d$ boundary is infinitely extended. In this sense we can
consider field theory on this $AdS_5$
at scales much larger than $L$. For exactly the same reason as
in the case of SYM theory, all we would see at low energies
should be the projection
of the $5d$ world onto its own flat boundary. Can we
identify this projection with the observable flat $4d$ world?

\vskip3mm
\noindent
\subsubsection*{2. Hidden cosmological constant}\scs{2}

\vskip2mm
\noindent
Before discussing some obvious objections against this
scenario, let us demonstrate in more detail
why in this setup the cosmological
constant $\Lambda_4$ of the 4--dimensional world can be zero: 
the 5--dimensional cosmological constant $\Lambda_5$ 
does not induce a 4--dimensional cosmological constant but gets
absorbed in the $AdS_5$ curvature radius. 
This is essentially
a new version of a mechanism discussed long ago by Rubakov and Shaposhnikov
\cite{rub}.\footnote{as the author found out after this paper was written.}
Very similar suggestions by E. and H. Verlinde 
have also just appeared in \cite{ver2}.

Let us first consider Einstein gravity.
Also, we start with the case where the $4d$ metric
 $g_{\mu\nu}=e^{2\alpha(\phi)}\hat g_{\mu\nu}$ has only an overall
$\phi-$dependence:
$$ds^2=d\phi^2+e^{2\alpha(\phi)}\hat g_{\mu\nu}(x)dx^\mu dx^\nu\ .$$
$\hat g$ is some background metric 
which for now we allow to be non--flat.\footnote{In general, 
$g(x,\phi)$ will not factorize into a $\phi$--dependent piece
and an $x$--dependent piece. In this case one should define $\alpha$ via
the logarithm of the $4d$ volume:
$e^{4\alpha(\phi)} \equiv \{\int d^4x\ {\sqrt{g^{(4)}(x,\phi)}}\}/
\{\int d^4x\}\ .$}

The Einstein equations imply (we redefine $\Lambda_5$ by a factor)
\ba R_{\mu\nu}^{(5)}\ =\ \ g_{\mu\nu}\ \Lambda_5
\la{maja}\ea
with $\Lambda_5$ assumed negative.
The LHS
of (\ref{maja}) can be written in terms of the $4d$ curvature tensor ${\hat R}^{(4)}$
of the metric $\hat g$ and the ``shifted dilaton'' (a slight misnomer here)
$$\varphi \equiv - \log({\sqrt g})\ =\ -\ 4\alpha(\phi)\ +\ \hbox{const.} $$
in the form
\ba
R_{\mu\nu}^{(5)}\ =\ 
  {\hat R}^{(4)}_{\mu\nu}\ +\ {1\over4}\ e^{-{\varphi\over2}}{\hat g}_{\mu\nu}
(\ddot\varphi\ -\ \dot\varphi^2)\ ,\la{box}\ea
where ``dot'' means ``$d/d\phi$''.
From this and (\ref{maja}) we read off that $\hat g$ is an Einstein metric
with effective $4d$ cosmological constant $\Lambda_4$ (also redefined by a factor):
\ba{\hat R}_{\mu\nu}^{(4)}\ =\ {\hat g}_{\mu\nu}^{(4)}\ \Lambda_4\la{mix}\ ,\ea
where
\ba\Lambda_4\ =\ e^{-{\varphi\over2}}[\Lambda_5\ -\ {1\over4}(\ddot\varphi-\dot\varphi^2)]\ \la{1}\ea
is $\phi$--independent.

$\ddot\varphi$ can be eliminated using the $g_{\phi\phi}$--equation of motion
\ba\ddot\varphi-{1\over4}\dot\varphi^2=\Lambda_5\ .\la{2}\ea
The result is
\ba\Lambda_4\ =\ {3\over4}e^{-{\varphi\over2}}(\Lambda_5+{1\over4}\dot\varphi^2)\ .\la{3}\ea
We see that there is a solution in which the $5d$ cosmological constant
is completely cancelled by $\dot\varphi^2$, and the
$4d$ space is flat:
$$\dot\varphi^2\ =\ -\ 4\Lambda_5\ =\ \ \ \rightarrow\ \ \ 
\Lambda_4=0\ .$$
This is $AdS_5$ (recall that $\Lambda_5$ is negative).
Of course there are also classical solutions $\varphi(\phi)$
with $\Lambda_4\neq0$. Those just
correspond to different foliations of
$AdS_5$ by $4d$ hypersurfaces, so in the present case the value of $\Lambda_4$
is in fact ambiguous.
But in the next section, a particular foliation will be singled out
such that $\Lambda_4$ is well--defined.

\vskip3mm
\noindent
\subsubsection*{3. Breaking conformal invariance}\scs{3}

\vskip2mm
\noindent
Let us now come to two immediate objections against
the proposal that our world is the projection of a 5--dimensional anti-de-Sitter world
onto its flat boundary:
the 4--dimensional world that this scenario
 predicts differs from the one we observe in at least two major aspects: 
\begin{enumerate}
\item
it is conformally invariant.
\item
it has zero 4--dimensional Newton constant.
\end{enumerate}
The first point reflects the fact that the SO(2,4) symmetry group of $AdS_5$
becomes the conformal group on the $4d$ boundary.
To break conformal invariance, we can let
the $5d$ geometry deviate from $AdS_5$ near its boundary.
Such a deviation would seem to be natural for a physical system with
a ``surface'', such as the $5d$ universe with $4d$ boundary.\footnote{This scenario
of a $4d$ universe whose conformal invariance is slightly broken 
might be related to the proposal \cite{fram} of conformal invariance at the TeV scale.}

More precisely, we will assume that the $5d$ cosmological constant
becomes $\phi$--dependent near the $AdS$--boundary:
$$\Lambda_5(\phi) .$$
Note that this implicitly singles out a particular foliation of $AdS_5$
by 4--dimensional hypersurfaces: those of constant $\Lambda_5$.
A $\phi$--dependent $\Lambda_5$ can arise
when $5d$ Einstein gravity is embedded 
in string theory, in particular on perturbations of the
 $AdS_5\times E^5$ backgrounds mentioned in the introduction.\footnote{In this way
one can also add all kinds of $4d$ matter and
gauge fields to the scenario, in terms of
Kaluza--Klein modes on $E^5$ or branes wrapped around cycles in $E^5$.}
At low energies this corresponds to embedding Einstein gravity in
 $5d$ gauged supergravity. 
There, $\Lambda_5$ is related to the potential $V[\Phi^I(\phi)]$,
where $\Phi^I$ are the scalar fields of gauged supergravity. Those
fields in turn generally depend on $\phi$. The way in which they roll  down the
potential $V(\Phi)$ as a function of $\phi$
 then encodes, i.p., the details of the breakdown
of the conformal and other symmetries. 

E.g., the $5d$ geometry might be the holographic dual of RG flows in
$4d$ gauge theories that approach a fixed point in the IR (the interior
of $AdS_5$) but not in the UV (the boundary of $AdS_5$).
Explicit examples of $5d$ geometries with varying potential $V$ 
have been discussed e.g. in \cite{porr,war}; although most of the
examples correspond, in the dual picture, to
 RG flows between UV and IR  fixed points, there are also flows without 
UV fixed points.

Assuming a constant dilaton $\Phi$\footnote{which is the case
in the solutions in \cite{porr,war}; the general case would require
redefining  $\varphi = 2\Phi - \log{\sqrt{g}} $
and also involves the dilaton equation of motion.},
the equations of motion of section 2 including (\ref{3}) remain the same 
even in the case of a $\phi$--dependent $\Lambda_5$,
and the conclusion
is also the same:
there exists a solution for $\varphi(\phi)$ with
$$\dot\varphi^2=-4\Lambda_5(\phi)\ \ \ \rightarrow\ \ \ \Lambda_4=0$$ 
everywhere, despite of the fact that conformal symmetry is broken. 
The general solution for $\varphi(\phi)$ has a constant but nonzero
$\Lambda_4$ in (\ref{3}). $\Lambda_4$ is $\phi$--independent
 since $\hat g$ and $\hat R$
  are $\phi$--independent in (\ref{mix}).
But if $\Lambda_4$ is zero at one $\phi$,
it is zero for all $\phi$. 

So we only need to argue that $\Lambda_4$ is zero in the interior of $AdS_5$.
At this point it might seem that we have merely replaced 
one fine--tuning problem
by another one: now we have to fine--tune $\dot\varphi^2$ to
exactly cancel the five--dimensional cosmological constant.

However, the situation does seem to have improved. Before, we had
to fine-tune the $4d$ cosmological constant at high energies to
a precise {\it nonzero} value, such that the low--energy cosmological constant
ends up being exactly zero, which seemed absurd.
Now, by contrast, we only have to find an argument why 
the hypersurfaces of constant $\Lambda_5$
 defined by the $5d$ gauged supergravity solution
should be flat. This seems much more natural.
We will return to this issue in the future.

So here we have a ``holographic mechanism''
that keeps the visible 4--dimensional cosmological
constant zero with the help of
one extra dimension:
 the 5--dimensional cosmological constant $\Lambda_5$ 
may be huge and moreover may change as a function of $\phi$,
as matter fields roll down some potential $V$ (as in the above
example of $5d$ gauged supergravity). But 
$\Lambda_4$ remains zero:
$\Lambda_5(\phi)$ is completely
absorbed by the $\phi$--dependence of $\varphi$ (i.e. of the ``warp factor'').

\vskip3mm
\noindent
\subsubsection*{4. Adding $4d$ gravity}\scs{4}

\vskip2mm
\noindent
The second immediate objection mentioned above was that the $4d$ world that
lives on the AdS--boundary has zero Newton constant.
A related point is that the modes of the $5d$ graviton that 
would represent massless $4d$
gravitons are not normalizable: there is no dynamical $4d$ gravity.\footnote{I thank
P. Horava for first making this point.}
To recall why the $4d$ Newton constant is zero
we use the form
$$ds^2\ =\ {L^2\over z^2}(dz^2+dx_{||}^2)$$
of the $AdS$--metric.
As in ordinary Kaluza--Klein compactification on a circle (instead of the
noncompact coordinate $z$),
integrating the $5d$ Hilbert-Einstein action over the coordinate $z$
 yields a $4d$ Hilbert-Einstein action:
$${1\over\kappa^2_5}\int dz\ d^4x_{||}\ {\sqrt{g^{(5)}}}R^{(5)} 
\ \ \rightarrow\ \ {1\over\kappa_4^2}\int d^4x_{||}\ {\sqrt{{\hat g}^{(4)}}}{\hat R}^{(4)}$$
with
$$\ {1\over\kappa_4^2}\ =\ {1\over\kappa^2_5}
\int_0^\infty {L^3\over z^3}\ dz\  .$$
The problem is that this integral diverges and therefore
$\kappa_4$ is zero.

This point may also be overcome by letting the geometry deviate
from $AdS_5$ near its boundary at $z=0$. A radical deviation would be to
simply cut off
the $5d$ universe near its boundary following Randall and Sundrum \cite{rs}:
we restrict $z$ to the range
$$z\ge\epsilon\ .$$
This corresponds to an explicit sharp UV cutoff in the dual gauge theory.
Now there are propagating $4d$ gravitons and the $4d$ Newton constant
is nonzero:
$${\kappa_4^2}\ =\ {1\over3}{\kappa^2_5}\ {\epsilon^2\over L^3}\ .$$

Of course, in general the value of the $4d$ Newton constant is 
not universal: it will depend on how precisely $AdS_5$
is cut off near its boundary. E.g., 
instead of a sharp cutoff one might try to look for smooth modifications of
the metric near the $AdS$--boundary such that the warp factor,
instead of being proportional to ${1\over z^2}$, converges 
to zero at $z=0$.
This would also lead to a non--vanishing Newton constant
and to dynamical $4d$ gravity.
One might ask at first what kind of gauge theory flow such a 
geometry would be dual to.
But the presence of $4d$ gravity seems to suggest
that such geometries have no pure gauge theory interpretation;
rather, the gauge theory should be embedded in a string theory.
In fact, if the warp factor goes to zero at the boundary,
the metric $g_{\mu\nu}$ goes to zero. This suggests that one should 
perhaps think of the radial direction of $AdS_5$ as being compactified.
In building a consistent compactification, presumably one then
inevitably ends up with the type of string compactification
studied in \cite{ver1}.

At this point the simple picture we started with becomes complicated.
In particular, it seems to be no longer clear how to argue
that the vanishing of $\Lambda_4$ inside $AdS_5$ implies 
the vanishing of $\Lambda_4$
near the ``boundary''. Certainly the equations used in section 2 are
no longer appropriate -- there are $\alpha'$--corrections
and string loop corrections.\footnote{I
thank P. Mayr for critical questions in this context.}

In conclusion, the conjecture that our flat 4--dimensional
 world is only the flat
projection of a strongly curved 5--dimensional world seems intriguing,
but solving the cosmological constant problem requires,
at the least, further thought.

\subsection*{Acknowledgements}

I would like to acknowledge interesting discussions 
about previous drafts of this note with E. and H. Verlinde,
who have simultaneously pursued closely related ideas \cite{ver2}.
This work is supported by a Heisenberg fellowship of the DFG.

\newpage

\end{document}